\documentclass{article}

\usepackage{graphics, graphicx, comment}
\usepackage{color}
\usepackage{latexsym}
\usepackage{epsfig}
\usepackage{subfigure}
\usepackage[margin=2.5cm, centering]{geometry}

\graphicspath{./figures/}

\newtheorem{lemma}{Lemma}

\begin{document}

\title{Robust Computation of a Minimum Area Convex Polygon Stabber of a Set of Isothetic Line Segments}
\maketitle 

\begin{flushleft}
Xin Wu \\
School of Computer Science \\
University of Windsor, ON N9B 3P4 \\
Canada
\end{flushleft}

\begin{flushleft}
Xijie Zeng \\
School of Computer Science \\
University of Windsor, ON N9B 3P4 \\
Canada
\end{flushleft}

\begin{flushleft}
Bryan St. Amour \hfill stamoub@uwindsor.ca \\
School of Computer Science \\
University of Windsor, ON N9B 3P4 \\
Canada
\end{flushleft}

\begin{flushleft}
Asish Mukhopadhyay \hfill asishm@uwindsor.ca\\
School of Computer Science \\
University of Windsor, ON N9B 3P4 \\
Canada
\end{flushleft}

\hrule
\begin{center}
{\Large {\bf Abstract}}
\end{center}

\noindent
In this paper, we discuss the algorithm engineering aspects of an $O(n^2)$-time algorithm 
\cite{paper:mgr} for computing a minimum-area convex polygon that intersects a set of $n$ isothetic line segments. \\

\vspace{0.5cm}
\hrule

\maketitle

\section{Introduction}

Let $S = \{s_1, s_2, \ldots, s_n\}$ be a set of $n$ isothetic line segments in the plane. Let $P$ be any 
convex polygon. A segment $s_i$ intersects $P$ if it lies in the interior of $P$ or intersects its boundary.     
In \cite{paper:mgr}, Mukhopadhyay et. al. proposed an $O(n^2)$ algorithm for computing such a polygon of minimum area, 
$P_{min}$. \\

The algorithm engineering issues that arise in the implementation of a geometric algorithm can be quite 
challenging. We wanted to experience this first-hand and hence the ab initio effort of this paper towards a robust implementation of the above 
algorithm. Below, we discuss the challenges faced and the strategies we devised to overcome these.\\

To make the paper self-contained, we outline in the next section the salient features of the algorithm in 
\cite{paper:mgr}, along with details of how each of the major steps was implemented.
In the following section, we discuss our handling of the input and output.
Results of some experiments are provided in the next section, and finally we conclude.

\section{Algorithm outline}    
\begin{figure}[h!]
\centering
\includegraphics[scale=1.0]{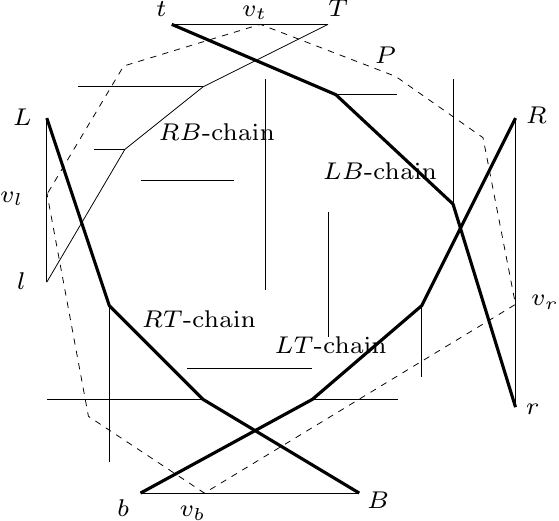}
\vspace*{8pt}
\caption{{\em Four hull chains for isothetic line segments \label{fig:pic1}}}
\end{figure}

A convex polygon is assumed to intersect a line segment when the latter lies entirely inside the polygon or intersects its boundary. 
As an example is a useful starting point in designing a geometric algorithm, we refer the reader to  
Fig.~\ref{fig:pic1}, where the dotted convex polygon
is a potential minimum area polygon that intersects all the segments. \\    

So is some 
terminology, that we introduce next. Four functions are associated with each line segment $s$: {\tt top}($s$), {\tt bot}($s$),
{\tt left}($s$) and {\tt right}($s$) that respectively returns the top, bottom, left and right end-point of $s$.
For a vertical line segment, the functions {\tt left}() and {\tt right}() are undefined,
while {\tt top}() and {\tt bot}() are undefined for a horizontal line segment. \\

{\bf Step 1: Finding extreme segments} \\

Four such segments in $S$ are found as below. For now, we assume that these exist and are unique, 
deferring the discussion of degenerate cases to a later section. 

\begin{itemize}

\item A ``top-most'' extreme segment, $\overline{tT}$: Find a vertical segment $t_1$ such that 
{\tt bot}($t_1$) $>$ {\tt bot}($s_i$) for all $s_i \in S$, and the highest horizontal segment $t_2$. 
If {\tt bot}($t_1$) is above $t_2$, then
$t_1$ is the ``top-most'' extreme segment. Otherwise, $t_2$ is the ``top-most".

\item A ``right-most'' extreme segment, $\overline{rR}$: Find a horizontal segment $r_1$ such that {\tt left}($r_1$) $>$ {\tt left}($s_i$) for all $s_i \in S$,
and find the vertical segment $r_2$ with the largest $x$-value. If {\tt left}($r_1$) is to the right of $r_2$,
then call $r_1$ the ``right-most'' extreme segment. Otherwise, $r_2$ is the ``right-most".

\item A ``bottom-most'' extreme segment, $\overline{bB}$: Find a vertical segment $b_1$ such that 
{\tt top}($b_1) < {\tt top}(s_i)$ for all $s_i \in S$, and the lowest horizontal segment $b_2$.   
If {\tt top}($b_1$) is below  $b_2$ then $b_1$ is the ``bottom-most" extreme segment. Otherwise, $b_2$ is the ``bottom-most".

\item A ``left-most'' extreme segment, $\overline{lL}$: Find a horizontal segment $l_1$, 
such {\tt right}($l_1$) $<$ {\tt right}($s_i$) for all $s_i \in S$, and find the vertical segment $l_2$ with the smallest $x$-value.  If {\tt right}($l_1$) is to the left of $l_2$,
then $l_1$ will be the ``left-most'' extreme segment. Otherwise, $l_2$ is the ``left-most" one.

\end{itemize}

In Fig.~\ref{fig:pic1}, the segments $\overline{tT}$ and $\overline{bB}$ are horizontal, while 
 $\overline{rR}$ and $\overline{lL}$ are vertical. For a given input, these extreme segments may not 
all exist and even when they
do there can be more than one extreme segment of a given type. We discuss later how our implementation deals with such 
degenerate input.   \\

{\bf Step 2: Computing critical chains} \\

Next we compute, using a standard convex hull algorithm \cite{book:ps}, the following 4 convex chains. 

\begin{itemize}
\item A convex chain, $RB$, going from $\overline{lL}$ to $\overline{tT}$ that is part of the convex hull of {\tt right}($s$) and {\tt bot}($s$)
of all segments $s$ in $S$, whenever these are defined.

\item A convex chain, $LB$, going from $\overline{tT}$ to $\overline{rR}$ that is part of the convex hull of {\tt left}($s$) and {\tt bot}($s$)
of all segments $s$ in $S$, whenever these are defined. 

\item A convex chain, $LT$, going from $\overline{rR}$ to $\overline{bB}$ that is part of the convex hull of {\tt left}($s$) and {\tt top}($s$)
of all segments $s$ in $S$, whenever these are defined. 

\item A convex chain, $RT$, going from $\overline{bB}$ to $\overline{lL}$ that is part of the convex hull of {\tt right}($s$) and {\tt top}($s$)
of all segments $s$ in $S$, whenever these are defined. 
\end{itemize}

These chains (refer again to Fig.~\ref{fig:pic5} for an example set of segments)  
help us  formulate a structural characterization of convex polygons that intersect all the segments 
in $S$ as in the lemma below.

\begin{lemma}
Let $P$ be any convex polygon that intersects all the line segments in $S$. Then the upper-left convex chain of $P$ 
must be on or above and to the left of the $RB$-chain; the upper-right chain of $P$ must be on or above and to the right of
the $LB$-chain; the lower-right chain of $P$ must be on or below and to the right of the $LT$-chain, and;
the lower-left chain of $P$ must be on or below and to the left of the $RT$-chain.
\end{lemma}

{\bf Proof:} See \cite{paper:mgr}. \hfill $\Box$ \\    

{\bf Step 3: Computing $P_{min}$} \\

It is possible that for a given set of input segments one or more of these chains may be missing. More about this
later. \\  

Thus, the class of convex polygons $P$ under consideration consist of \emph{at most} four convex chains, each of which 
joins a pair of extreme vertices ($v_l, v_t, v_r$, and $v_b$) that lie respectively on the extreme segments 
$\overline{lL}$, $\overline{rR}$, $\overline{tT}$, and $\overline{bB}$. \\

\begin{figure}[h!]
\centerline{\includegraphics[scale=0.6]{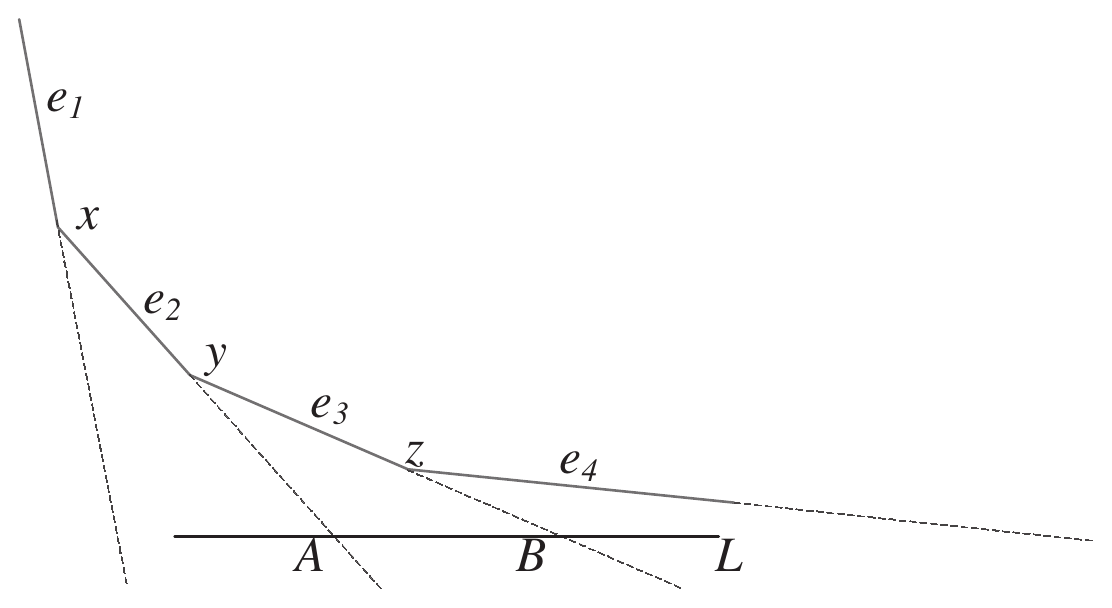}}
\vspace*{8pt}
\caption{{\em Partitioning an extreme segment into subintervals}}
\label{fig:wfigTwo}
\end{figure}

We can be more precise about the 
location of these extreme vertices. For example, $v_l$ will lie in one of the subintervals of $\overline{lL}$,
obtained by extending the edges of the critical chains to partition this extreme segment. The points in a 
given subinterval form an equivalence class in the sense that the edges of an intersecting convex polygon $P$ 
incident on a point of this subinterval are tangent to the critical chains at the same points. \\
In Fig.~\ref{fig:wfigTwo}, for example, the extreme segment $L$ is partitioned by the points $A$ and $B$ 
into three intervals. \\

Sub-interval generation was implemented by the method setIntervalsBy({\em MPCriticalChain}) of class 
MPExtremeSegment. The pseudo-code for this is shown below. \\ 

\newpage
\hrule
\vspace{0.5cm}
{\bf Algorithm} setIntervalsBy($MPCriticalChain$)

\begin{enumerate}
  \item Find the first edge $e$ of critical chain which can partition extreme segment.

  \item Initiate the tangent point of each interval to the first endpoint of $e$.

  \item For each edge $E$ of critical chain after $e$:

  \begin{description}
    \item{3.1.} Get the intersection point $P$ by extending edge $E$.

    \item{3.2.} If $P$ is null, continue.

    \item{3.3.} For each interval $I$ on extreme segment:

    \begin{description}
      \item{3.3.1.} If $P$ is located on interval $I$:
      \begin{description}
        \item{3.3.1.1.} If $P$ is identical with the first endpoint of interval $I$, set tangent point of $I$ to the second endpoint of $E$

        \item{3.3.1.2.} Else:
        \begin{description}
          \item{3.3.1.2.1.} Split interval $I$ into two parts, $I1$ and $I2$.

          \item{3.3.1.2.2.} Set tangent point of $I2$ to the second endpoint of $E$
        \end{description}

        \item{3.3.1.3.} Set each tangent point of intervals behind $I2$ to the second endpoint of $E$
      \end{description}
    \end{description}
  \end{description}
\end{enumerate}
\vspace{0.25cm}
\hrule
\vspace{0.5cm}

An input requirement of the above method is that all edges of a critical chain and the two endpoints of any edge be given in counterclockwise order. For the example in Fig.~\ref{fig:wfigTwo}, this method will set $x$ as the tangent point of the 
left interval on $L$, $y$ as the tangent point of the middle interval, and $z$ is the tangent point of the right interval. \\

In addition to generating intervals by extending edges of critical chains, intervals could also be created 
by the cutting of other extreme segments. Experiments show that this helps in determining $P_{min}$. 
The method setIntervalsBy({\em MPExtremeSegment}) perform this task. \\

\begin{figure}[tbp]
\centering
\includegraphics[scale=1.0]{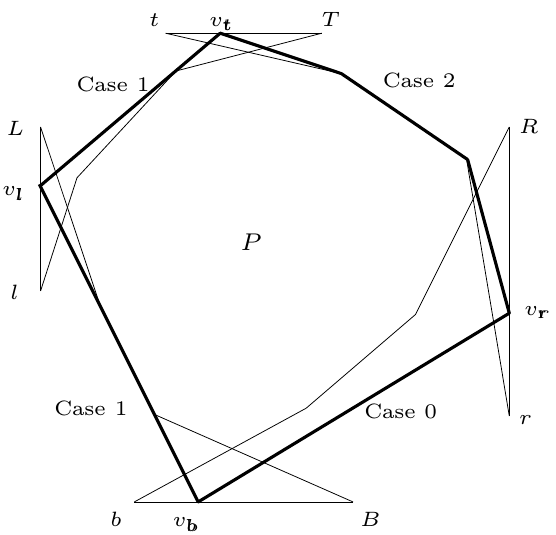}
\vspace*{8pt}
\caption{{\em An illustration of the different types of connection \label{fig:pic3}}}
\end{figure}
\begin{figure}[h!]
\centering
\subfigure[]{
\includegraphics[scale=1.0]{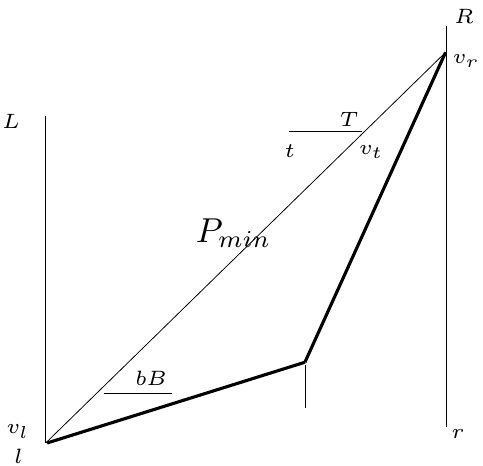}
\label{fig:pic4}
}
\quad
\subfigure[]{
\includegraphics[scale=1.0]{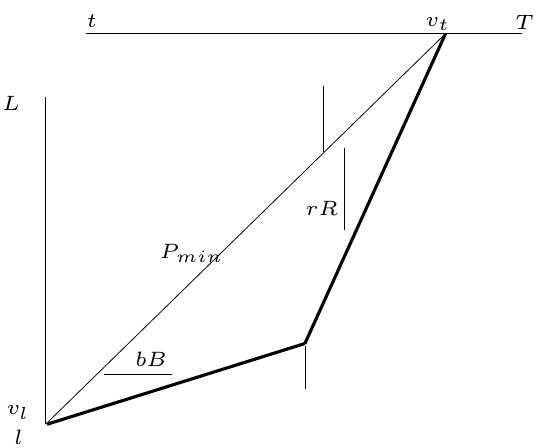}
\label{fig:pic5}
}
\caption{{\em Further subdivision of connection types:
          (a) The connection in bold bypasses $\overline{bB}$.
					(b) The connection in bold bypasses $\overline{rR}$ and $\overline{bB}$.
					}}
\label{fig:pic4and5}
\end{figure}

A convex chain of $P_{min}$ that joins two extreme vertices is defined as a \emph{connection}, and 
is one of the following types (see Fig.~\ref{fig:pic3}).

\begin{description}

\item[Case 0] A single edge that does not touch any of the critical chains.

\item[Case 1] A single edge that is tangent to an underlying critical chain.

\item[Case 2] A convex chain composed of multiple edges; in this case, an underlying critical chain
contributes some structure to it.

\end{description}

Two extreme vertices, one of which lies on a vertical (horizontal) extreme segment and 
the other on a horizontal (vertical) extreme segment
are considered to be \emph{adjacent}; otherwise, they are \emph{non-adjacent}.
Each of the above types of connections can be further subdivided according to which pairs of extreme
vertices they join. These are:

\begin{description}

\item[(i)] A connection joining two adjacent extreme vertices (for example: $v_l$ to $v_t$; $v_t$ to $v_r$; etc.;
see Figure~\ref{fig:pic3}).

\item[(ii)] A connection joining two non-adjacent extreme vertices ($v_l$ to $v_r$ or $v_t$ to $v_b$),
bypassing one of the extreme segments. As a result, this extreme segment is forced to lie in the
interior of $P_{min}$. See Figure~\ref{fig:pic4}, in which $v_t$ is not actually a vertex of $P_{min}$, 
and in which $v_b$ does not exist.
              

\item[(iii)] A connection that joins a pair of adjacent extreme vertices, 
bypassing the other extreme segments. As a result, the bypassed segments
are forced to lie in the interior of $P_{min}$. See Figure~\ref{fig:pic5}.

\end{description}

However, it is impossible for a connection to bypass three extreme segments. For a proof see \cite{paper:mgr}. \\

In \cite{paper:mgr} it was shown that to compute $P_{min}$ we have to consider 219 different configurations 
of connections, calculated as follows:

\begin{itemize}

\item 81 different possibilities when there are four connections,

\item 108 possibilities when there are three connections ($3^3$ times four, for each of the four extreme segments that can be bypassed),

\item 20 possibilities when we have one regular connection and one that bypasses two extreme segments
(four, for each position of the regular connection, times five, since the connections can be 0-2, 1-2, 2-0, 2-1, or 2-2), and

\item 10 possibilities when we have two connections each bypassing one extreme segment
(two possibilities for the bypassed segments, times five, since the connections can be 0-2, 1-2, 2-0, 2-1, or 2-2).

\end{itemize}

{\bf Remark:} Interpret notation $m-n$ as: one chain is of Type $m$ and the other of Type $n$. \\

In our implementation we reduced this number to 47 only by considering isomorphism 
and symmetry among these different configurations. The details are as below \\

We use the notation $x_1x_2x_3x_4$ to denote a four
connection configuration where $x_i (i = 1, 2, 3, 4)$ is one of the three connection types: 0, 1 or 2;
$x_1$ connects $\overline{tT}$ and $\overline{lL}$;
$x_2$ connects $\overline{lL}$ and $\overline{bB}$; $x_3$ connects $\overline{bB}$ and $\overline{rR}$; $x_4$ connects $\overline{rR}$ and $\overline{tT}$. \\


In terms of this notation, $x_1x_2x_3x_4, x_2x_3x_4x_1, x_3x_4x_1x_2$ and $x_4x_1x_2x_3$ are all
isomorphic configurations that can be
handled by the same procedure by permuting the input parameters. For example, if the procedure
\emph{Comp1012($\overline{tT}$, $\overline{lL}$, $\overline{bB}$, $\overline{rR}$, RB, RT, LT, LB)} computes the minimum area polygon for the configuration 1012,
then \emph{Comp1012($\overline{lL}$, $\overline{bB}$, $\overline{rR}$, $\overline{tT}$, RT, LT, LB, RB)} will be invoked for the configuration 0121.
Similarly, by a  proper ordering of the parameters, \emph{Comp1012} can also
compute minimum area polygon for the configurations 1210 and 2101. \\

If we assume that the inputs are in the first quadrant of an orthogonal $xy$-coordinate system,
we will consider the configurations $x_1x_2x_3x_4$ and $x_4x_3x_2x_1$ as mirror images of each other with respect to
the $y$-axis. We will say such configurations are symmetric; they  can be handled by the same procedure. Suppose
the procedure \emph{Comp0012} computes the minimum area polygon for the configuration 0012. If we mirror all the inputs
with respect to the $y$-axis, call \emph{Comp0012} on the mirrored inputs and reflect the output in the $y$-axis,
we will have solved the problem for the configuration 2100 on the original
input segments. \\


We do not have to consider symmetries of $x_1x_2x_3x_4$ with respect to the  $x$-axis or the origin. This is because
$x_2x_1x_4x_3$ ($x$-axis symmetric configuration) and $x_4x_3x_2x_1$ are isomorphic configurations,
and so are $x_3x_4x_1x_2$ (origin symmetric) and $x_1x_2x_3x_4$. \\

By using these two optimization strategies, we get the following 47 non-isomorphic, non-symmetrical configurations.

\begin{itemize}
\item For all 4-connection configurations, there are 24 non-isomorphic ones in Table~\ref{tab:tab1}.
However, 0012 and 1200 are isomorphic pairs, 1200 and 0021 are symmetrical pairs. Thus, 0012 and 0021 can be handled by the same procedure.
The same reason applies to 0122 and 0221, 1120 and 1102. Consequently, there are 21 non-isomorphic, non-symmetrical 4-connection
configurations.

\begin{table}[h!]
\begin{center}
\begin{tabular}{|c|c|c|c|c|c|}
\hline  0000& 1111& 2222& 0111& 0222& 1222\cr \hline 1100& 2200&
1122& 0101& 0202& 1212\cr \hline 0001& 0002& 1112& 0012& 1012&
0122\cr \hline 1020& 1120& 1202& 0021& 1102& 0221\cr \hline
\end{tabular}
\end{center}
\caption{Non-isomorphic 4-connection configurations
\label{tab:tab1}}
\end{table}

\item For all 3-connection configurations, there are 27 non-isomorphic ones in Table~\ref{tab:tab2}.
(3 connections and 3 types for each connection; There is no need to multiply by four, because of isomorphism.)
Among these 27 non-isomorphic configurations, there are 18 non-symmetrical ones in Table~\ref{tab:tab3}.
Note that for a given triple $xyz$, it could be $x$, $y$, or $z$ that bypasses a segment.
That is why there appear to be duplicates in the table; they are necessary.

\begin{table}[h!]
\begin{center}
\begin{tabular}{|c|c|c|c|c|c|}
\hline  000& 010& 020& 001& 011& 021\cr \hline 002& 012& 022& 100&
110& 120\cr \hline 101& 111& 121& 102& 112& 122\cr \hline 200&
210& 220& 201& 211& 221\cr \hline 202& 212& 222& & & \cr \hline
\end{tabular}
\end{center}
\caption{Non-isomorphic 3-connection configurations
\label{tab:tab2}}
\end{table}

\begin{table}[b!]
\begin{center}
\begin{tabular}{|c|c|c|c|c|c|}
\hline  000& 010& 020& 001& 011& 021\cr \hline 002& 012& 022& 101&
111& 121\cr \hline 102& 112& 122& 202& 212& 222\cr \hline
\end{tabular}
\end{center}
\caption{Non-isomorphic non-symmetric 3-connection configurations
\label{tab:tab3}}
\end{table}

\item For all 2-connection (one regular, one bypasses two extreme segments) configurations, there are 5 non-isomorphic,
non-symmetrical ones 02, 12, 20, 21, 22. For configuration 02, Case 0 is a regular connection while Case 2 bypasses two
extreme segments. For configuration 20, Case 2 is regular, while Case 0 bypasses two. So 02 and 20 are a
non-isomorphic, non-symmetrical pair. The same reason applies to 12 and 21.

\item For all 2-connection (both bypass one extreme segment) configurations, there are 3 non-isomorphic, non-symmetrical ones 02, 12 and 22.

\end{itemize}

When considering a potential solution polygon consisting of at least three connections, we 
can discern one or more of the following patterns in the configuration of connections. \\ 

\begin{figure}[h!]
\centering
\includegraphics[scale = 0.6]{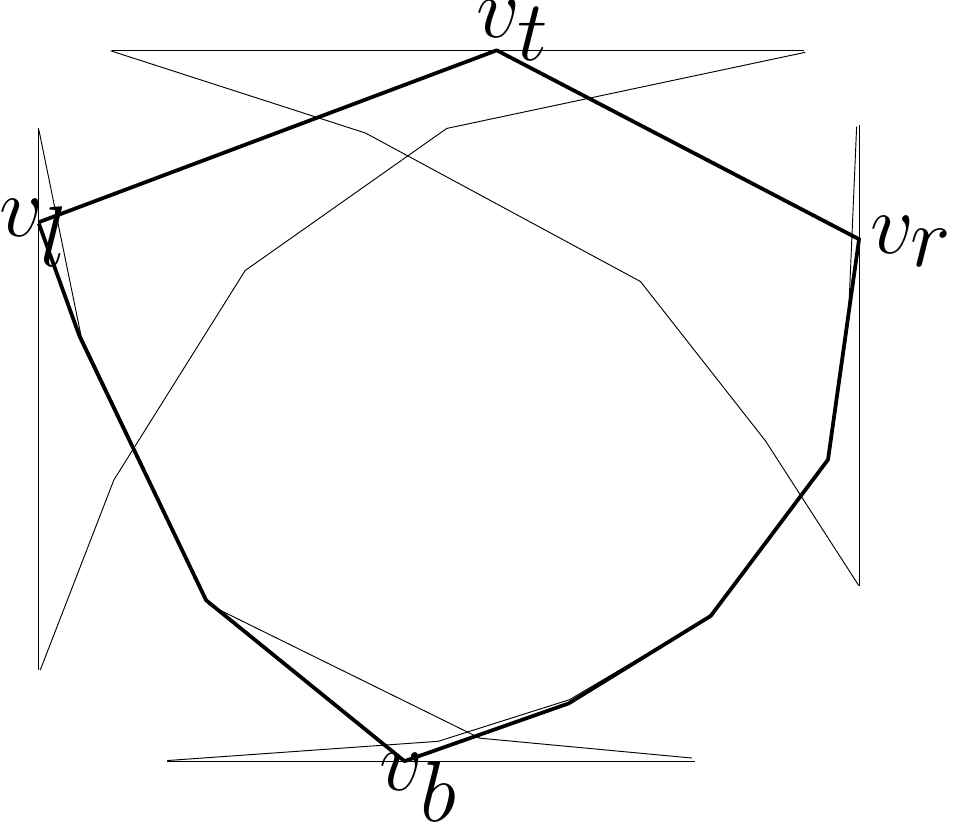}
\vspace*{8pt}
\caption{{\em Pattern A (lower half of bold polygon) and Pattern C (upper half of bold polygon) \label{fig:pic6}}}
\end{figure}

\begin{description}

\item[Pattern A] Two occurrences of Case 2: This divides the problem into two {\em independent} sub-problems.
If the two connections occur on ``adjacent'' critical chains (like $RB$ and $LB$, or $LB$ and $LT$, etc.) then
the problem is reduced to searching through $O(n)$ intervals on one extreme segment,
and searching through $O(n^3)$ interval triplets, for the other three extreme segments.
See Figure~\ref{fig:pic6} for an example of this.

If the connections occur on ``opposite'' chains ($RB$ and $LT$, or $RT$ and $LB$) then
the problem is reduced to choosing from $O(n^2)$ interval pairs for the two extreme segments on one side of
the chains, and choosing from $O(n^2)$ interval pairs for the two extreme segments on the other side.

\item[Pattern B] An occurrence of Case 1: (For example, the connection between $v_l$ and $v_t$ in
Figure~\ref{fig:pic3}.)
There are only $O(n)$ interval pairs that are connected by a line that is tangent to the underlying
critical chain. One can think of a tangent line rotating along the underlying critical chain:
this line will hit only $O(n)$ interval pairs. So, there will be $O(n^3)$ interval quadruplets to consider.

\item[Pattern C] Two adjacent occurrences of Case 0: (See Fig.~\ref{fig:pic6}) The extreme segment attached to these
connections
will not have to be divided into any intervals, since the underlying critical chains will not contribute
any structure to that part of $P_{min}$. Again, there will be only $O(n^3)$ interval quadruplets to consider.

\end{description}

These patterns can be used to show that only $O(n^2)$ different polygons have to considered to determine the minimum 
(see \cite{paper:mgr} for details). \\

Determination of the types of connections was an important part of our implementation. We describe below how
we went about this. \\  

The essence of this algorithm is to generate convex polygons by combining every possible connection between
extreme segments. It will search 47 possible configurations of connections in total. This implies the necessity of
determining whether a specific connection is Case 0, or Case 1, or Case 2, which are defined in 
section 2 and illustrated in Fig.~\ref{fig:pic3}. \\

Some involved functions are characterized as follows. \\

\emph{isCase0}

This function is used to check whether a connection between two adjacent extreme segments has the type of Case 0.
It returns true as long as such a connection has no intersection or overlapping with the corresponding critical chain
between same extreme segments.\\                                                                      

\emph{isCase0CommEndpoint}

According to the discussion of Pattern C, we know that when two adjacent connections are Case 0,
the common vertex of the two connections must be one of the endpoints of the corresponding extreme segment.
This also means that the common vertex is on a critical chain. This function is similar to isCase0,
but it allows one vertex of the connection to be on a critical chain instead of totally no touching. \\

\emph{isCase0Bypass}

A connection can join two non-adjacent extreme vertices, hence it bypasses one of the extreme segments. Consequently,
this extreme segment is forced to lie in the interior of the minimal convex polygon. Though there is no accurate
definition of Case 0 in bypass situation, this task can be performed in a somewhat tricky way. \\

In our implementation, a connection is considered having a type of Case 0 if

\begin{enumerate}

\item it touches the bypassed extreme segment, or it makes such extreme segment contained in the convex polygon; and

\item it is away from involved critical chains with the only exception that intersecting the nearest edge of
critical chain.

\end{enumerate}

The above two conditions make sure that the connection will not cut through the core area and, at the same time,
the candidate polygon will intersect the bypassed extreme segment. \\

\begin{figure}[h!]
\centerline{\includegraphics[scale=1.0]{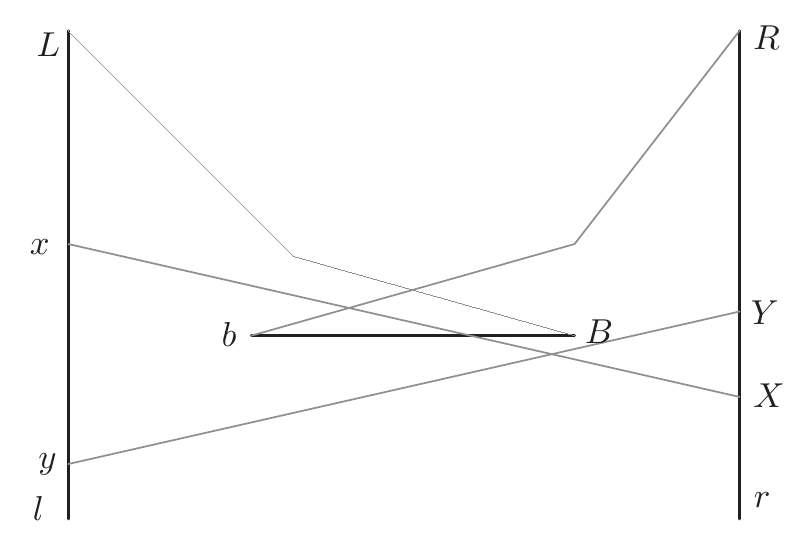}}
\vspace*{8pt}
\caption{{\em Two connections are Case 0 bypass \label{fig:wfig3}}}
\end{figure}

Two examples of Case 0 connection are illustrated in Figure~\ref{fig:wfig3}. Connection $\overline{xX}$ 
intersects extreme segment $\overline{bB}$, and does not cut through the core area which is partially 
bounded by the $RT$-chain and $LT$-chain. Undoubtedly, connection $\overline{xX}$ has the type of Case 0 
bypass. On the other hand, although connection $\overline{yY}$ does not intersect $\overline{bB}$, it is 
still capable of including $\overline{bB}$ in the polygon, to which $\overline{yY}$ contributes itself as
an edge. The consequential question is how to check the inclusion. Assuming $\overline{yY}$ is a directed 
line segment and follows the counterclockwise order, we can find that $\overline{bB}$ is included in the 
polygon if it is located on the left side of $\overline{yY}$. This is the exact approach that we implement in the program. \\


\emph{isCase0BypassCommEndpoint}

This function is used in the Pattern C when one of the two adjacent Case 0 connections is trying to bypass 
any extreme segment.  It is similar with isCase0Bypass but allows a common endpoint. \\

\emph{isCase1Bypass}

This function takes three points to represent a connection, that is, two endpoints and an in-between point 
at which the connection is tangential to the critical chain. Only by doing this we can make sure that, in a bypass situation, no extreme segment will be missed when generating polygons. \\

\emph{isCase2Bypass}

\begin{figure}[h!]
\centerline{\includegraphics[scale=0.8]{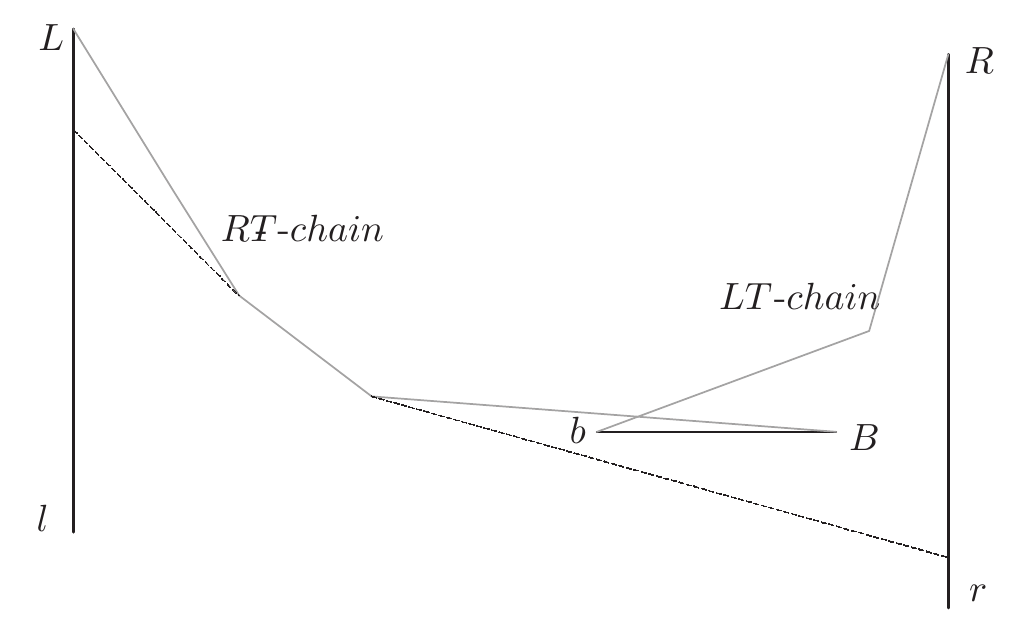}}
\vspace*{8pt}
\caption{{\em Case 2 bypass \label{fig:wfig4}}}
\end{figure}

This is similar to isCase1Bypass but it deals with the configuration illustrated in Figure~\ref{fig:wfig4}.

Here's a summary of the algorithm \\ 

\hrule
\vspace{0.5cm}
{\bf Algorithm} IsotheticMinPolyStabber($S$)

\begin{enumerate}
  \item Compute the critical chains $RB$, $LB$, $LT$, and $RT$.

  \item Extend the edges of these chains to partition the extreme segments $\overline{lL}$, $\overline{tT}$,
    $\overline{rR}$, and $\overline{bB}$;
    store the extended edges and corresponding points of tangency, on the critical chains,
    for each partition point.

  \item For each configuration of connections:

  \begin{description}
    \item{3.1.} For each possible tuple of intervals:

    \begin{description}
      \item{3.1.1.} Solve an optimization problem with suitable constraints, resulting in
        two to four extreme vertices.

      \item{3.1.2.} Find the area of the polygon using these extreme vertices.

      \item{3.1.3.} If this is the smallest polygon seen so far then store these extreme vertices
        as the optimal ones.
    \end{description}
  \end{description}

  \item Report the minimum polygon by joining the optimal extreme vertices, using their points of
    tangency to the critical chains.

\end{enumerate}
\vspace{0.25cm}
\hrule
\vspace{0.25cm}



\begin{figure}[h!]
\centerline{\includegraphics[scale=0.6]{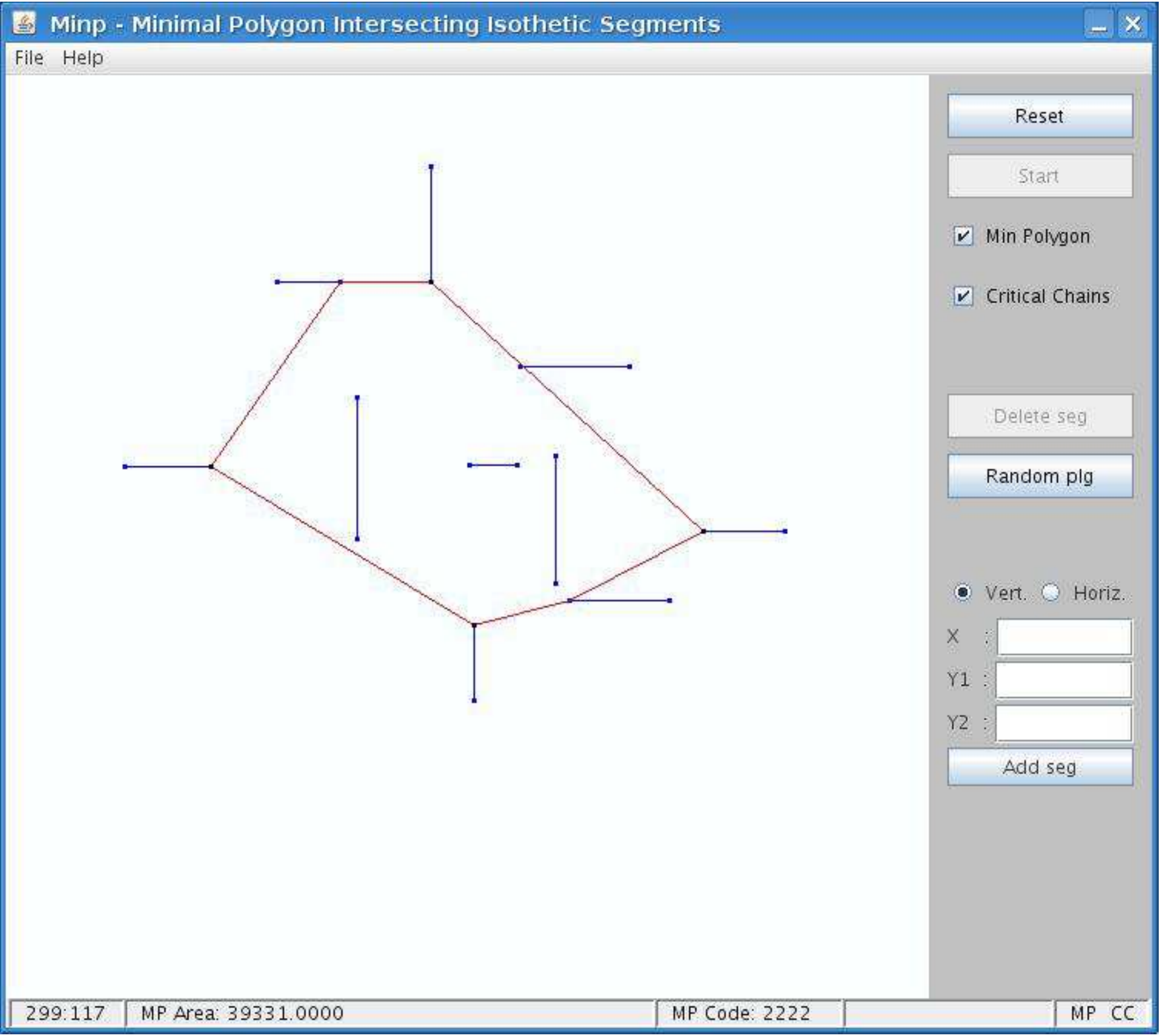}}
\vspace*{8pt}
\caption{{\em Screen shot of our application\label{fig:wfig2}}}
\end{figure}




\newpage
\section{Input/output handling} 

A given set of input segments can be degenerate in a variety of ways, as discussed below. In our 
implementation, we have attempted to handle this as exhaustively as possible. \\   

{\bf Line Stabbing} \\

A line stabber is a line that intersects all the  
segments in $S$. When such a line exists, $P_{min}$ reduces to a line segment of zero area. In this
case, our implementation simply reports the situation without producing  any output.\\

Our implementation detects the existence of a line stabber, based on the characterization in the following lemma, proved in \cite{paper:mgr}.

\begin{lemma}\label{lem:line_stab}
$l$ is a line transversal of the set of isothetic line segments if and only if it is
either above the convex hull of the set $S_{RB} = \{ {\tt bot}(s_i), {\tt right}(s_i) | i=1,2,\ldots, n\}$
and below the convex hull of the set $S_{LT} = \{ {\tt top}(s_i), {\tt left}(s_i) | i=1,2,\ldots, n\}$,
or below the convex hull of the set $S_{RT} = \{ {\tt top}(s_i), {\tt right}(s_i) | i=1,2,\ldots, n\}$
and above the the convex hull of the set $S_{LB} = \{ {\tt bot}(s_i), {\tt left}(s_i) | i=1,2,\ldots, n\}$.
\end{lemma}

{\bf Insufficient extreme segments} \\

If the input has fewer than 3 extreme segments, the input is not processed.  \\

{\bf Multiple extreme segments} \\

While the algorithm assumed that these extreme segments are unique, input data may not satisfy this. Suppose there
are multiple horizontal segments that are topmost. We can deal with this situation as follows. Let $TH$ be the set of 
topmost horizontal segments. Let the left endpoint of each segment 
be marked RED and the right endpoint of each segment be marked BLUE. Pick the leftmost-BLUE point and the 
rightmost-RED point. If the rightmost-RED point is to the left of the leftmost-BLUE point the segments in $TH$ have a 
common intersection that goes from the rightmost-RED to the leftmost-BLUE point and we replace $TH$ by this single 
segment. Else, our solution for $P_{min}$ will have a top horizontal segment that goes from the leftmost-BLUE point to the 
rightmost-RED point. \\

The correctness of this is not hard to prove. If every pair of segments in $TH$ have a common intersection, then by 
Helly's theorem all the segments in $TH$ have a common intersection. Otherwise, there exists a pair 
that does not intersect and hence a pair whose left and right end-points are farthest apart. These correspond to the 
leftmost-BLUE and rightmost-RED points in the above algorithm.  \\

In our implementation it was very easy to identify the case where the input data contains multiple extreme segments.
Handling that situation, however, was not as simple as identifying it. If $TH$ does not contain a common intersection
then the $RB$ and $LB$ convex chains intersect one another over a line segment, not a single point. Let $TI$ be the
intersection of the two chains, meaning if a point $p$ is covered by $RB$ and is part of the structure of $LB$,
then $p$ is in $TI$. We then add the points in $TI$ to both chains, which, due to the definition of $TI$, will not
change the shape of either chain. By doing this, we can now safely set either the leftmost-BLUE point, or the
rightmost-RED point to be the unique extreme segment, and continue forward with the algorithm.

\begin{figure}[h!]
\centerline{\includegraphics[scale=0.6]{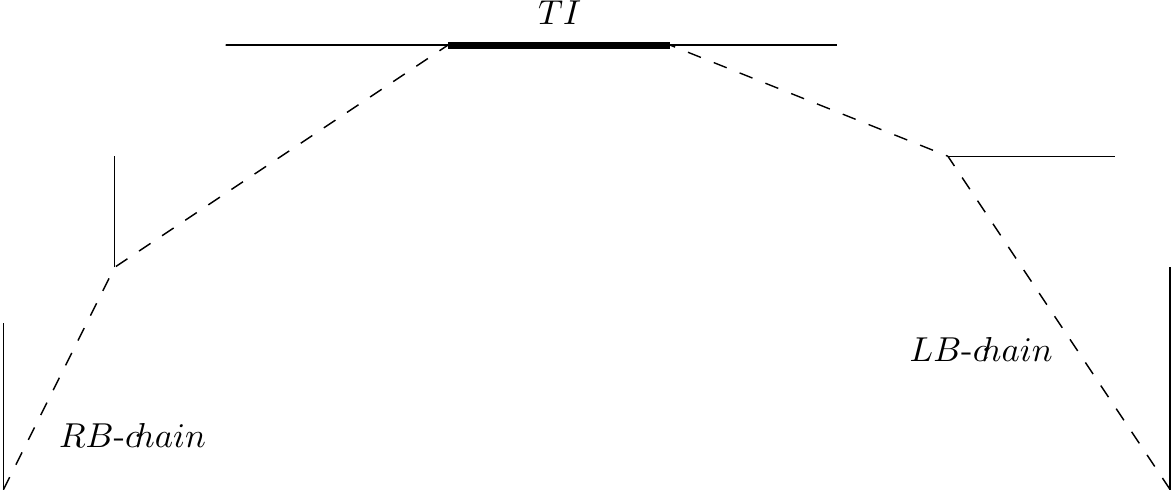}}
\vspace*{8pt}
\caption{{\em Finding the intersection of two chains\label{fig:wfig2}}}
\end{figure}
\quad\quad
       


{\bf Output handling} \\ 

The GUI displays $P_{min}$ in red and its area is shown in a text display bar at the bottom. An user can also verify the 
minimality
of the output polygon by sliding the extreme vertices on their respective segments when it is possible to do so. The text 
display shows the area of the altered polygon that can be checked against the area of $P_{min}$. The reader is invited to test 
the implementation by following the software link from the URL {\tt http://asishm.myweb.cs.uwindsor.ca}. At present, the applet runs 
only in the Firefox web-browser.


\section{Software Architecture}

The dependencies of the different packages are shown in the diagram below
This software largely uses inheritance, the main feature of OO-programming, to build up the complex functions.
The classes designed can be classified into four categories as follows.

\begin{enumerate}

\item GUI-purpose (package aljin.guilet): these classes fulfill the GUI interface such as menu bar and 
dialog windows,
deal with interactions between software and user. This package consists of various
event listeners and GUI components. For example, class MEasel provides an easy-to-use canvas; while class 
MArena is a graphic system to visualize user-defined geometrical objects and make them editable. 

\item CG-purpose (package aljin.cg): to provide implementations of objects with regard to computational geometry. Typical classes include CGPoint, CGLineSegment, and CGPolygon. 

\item Algorithm-purpose (package aljin.alg.minpolygon): implementation of algorithm. This package contains 
the core functions and some related algorithm-level classes.

\item Application-purpose (package default): these classes compose the executable application. The program 
entry is Minp.main().

\end{enumerate}

\begin{figure}[h!]
\centerline{\includegraphics[scale=0.8]{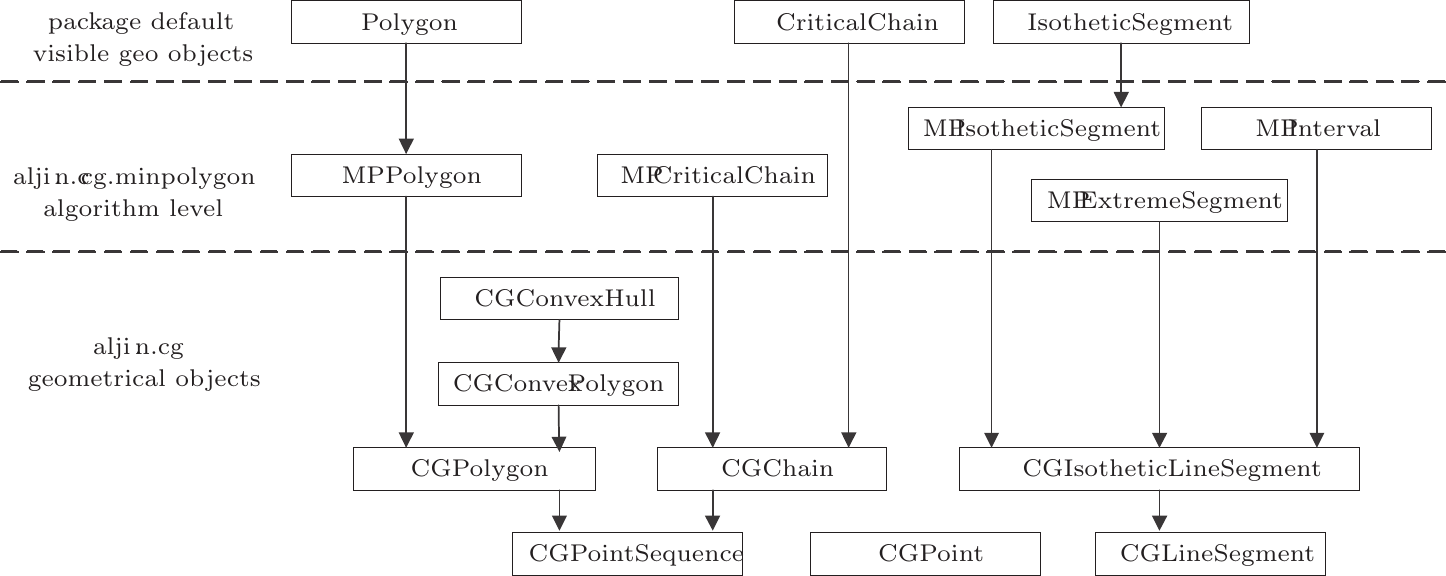}}
\vspace*{8pt}
\caption{{\em Hierarchy of geometry-related classes \label{fig:wfig5}}}
\end{figure}

Figure~\ref{fig:wfig5} illustrates the hierarchy of geometry-related classes, where solid directed segment 
means ``derived from''.

\section{Experimental Results}

In this section, we provide some experimental results to illustrate the two optimization technologies and the three
types of connections. \\

\begin{figure}[h!]
\centering
\subfigure[]{ 
\includegraphics[scale=0.6]{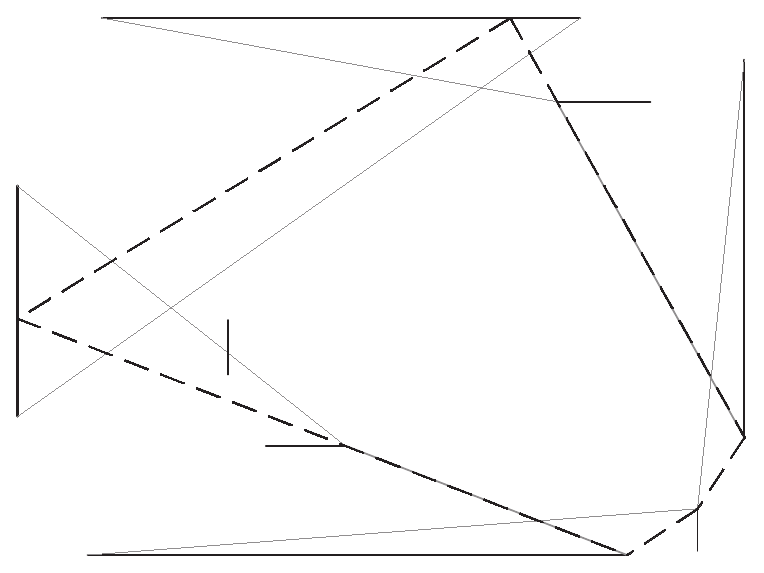} 
\label{fig12} 
}
\quad
\subfigure[]{
\includegraphics[scale=0.6]{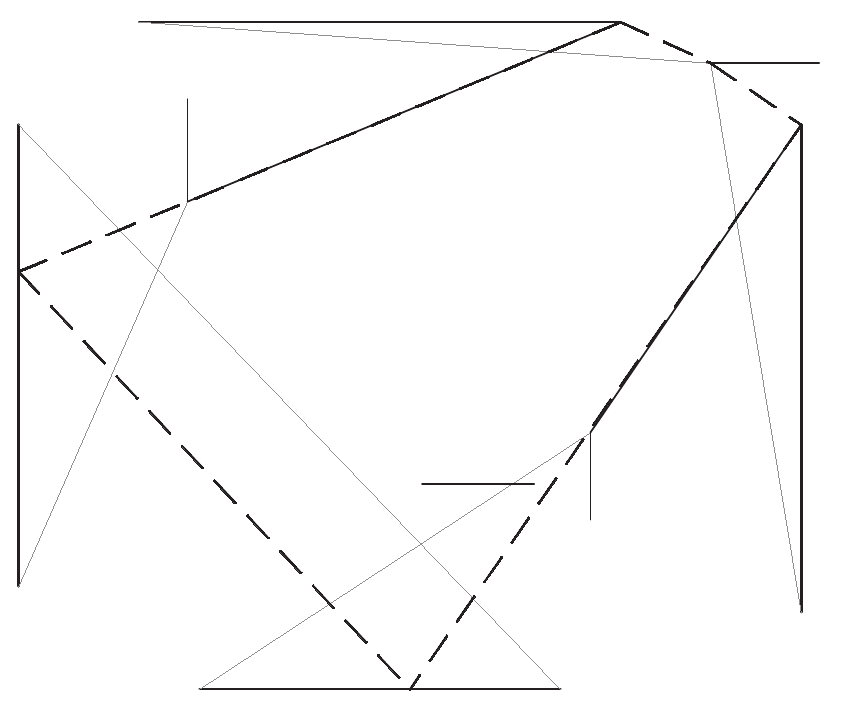} 
\label{fig13}
}
\quad
\caption{\em{Resulting polygon with (a) code 0121 (b) code 1012}}
\label{figs:12and13}
\end{figure}

In Figure~\ref{fig12}, the configuration of the minimal polygon is 0121 and, in Figure~\ref{fig13}, it is 1012. Since the input of the
former example is the outcome of 90 degree counterclockwise rotation of the later one, the application uses the
same procedure to compute the minimal polygon with different arguments. In fact, only the order of arguments is
different. Similarly, one procedure can be used to compute both 2101 and 1210. \\

\begin{figure}[h!]
\centering
\subfigure[]{ 
\includegraphics[scale=0.6]{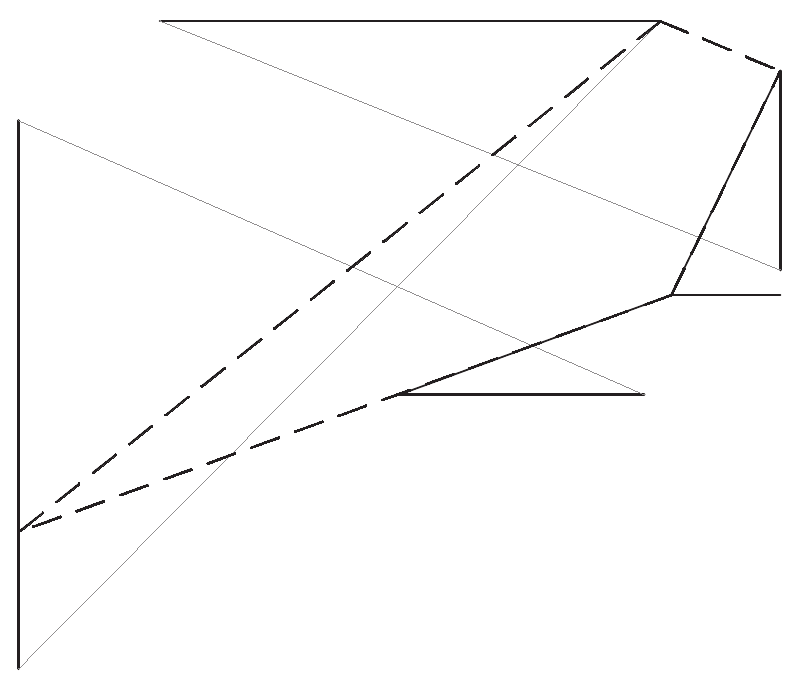} 
\label{fig14} 
}
\quad
\subfigure[]{
\includegraphics[scale=0.6]{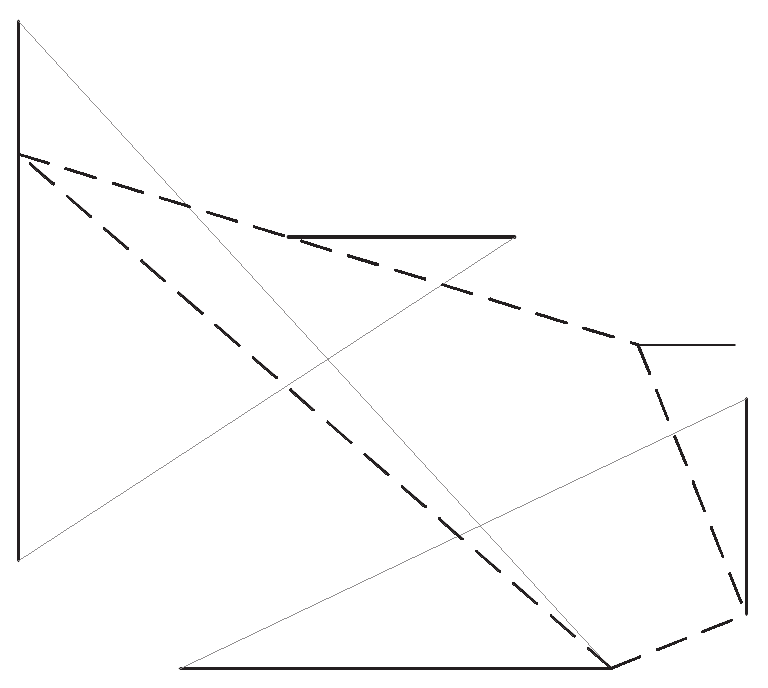}
\label{fig15}
}
\quad
\caption{\em{Resulting polygon with (a) code 120 (b) code 102}}
\label{figs:14and15}
\end{figure}

In Figure~\ref{fig14}, the configuration of the minimal polygon is 120 and, in Figure~\ref{fig15}, the result is 102.
Since the two inputs are $y$-coordinate symmetric, they are handled by the same procedure. \\



\begin{figure}[h!]
\centering
\subfigure[]{ 
\includegraphics[scale=0.6]{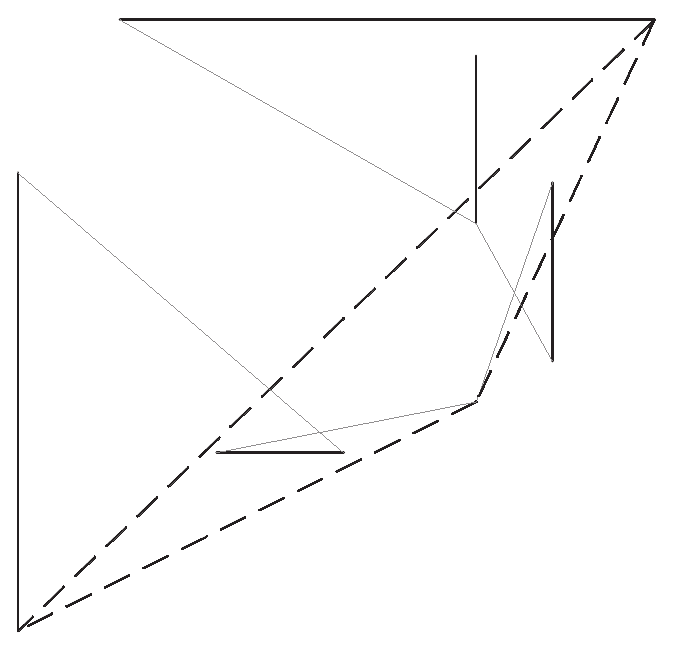} 
\label{fig16} 
}
\quad
\subfigure[]{
\includegraphics[scale=0.6]{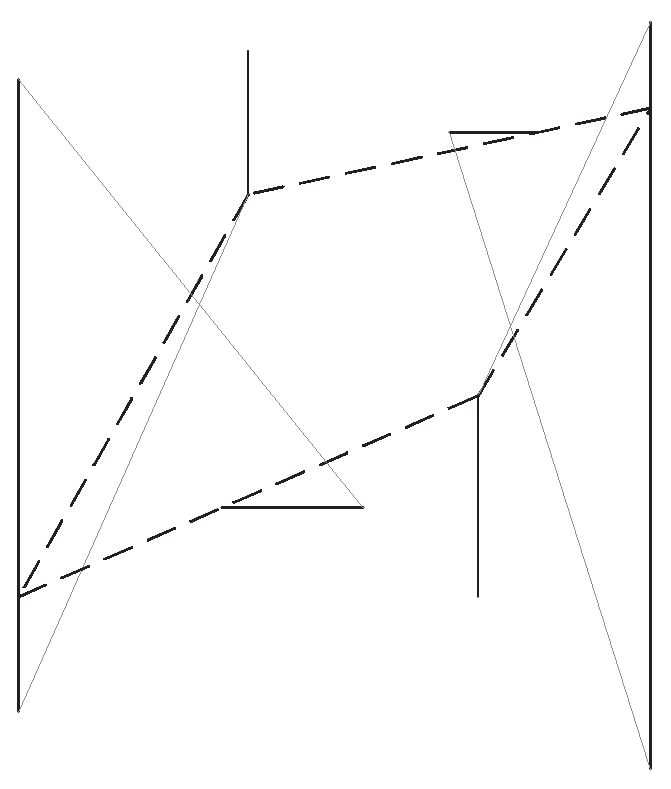}
\label{fig17}
}
\quad
\caption{\em{Resulting polygon with (a) code 12 (b) code 22}}
\label{figs:16and17}
\end{figure}

In Figure~\ref{fig16}, the configuration of the minimal polygon is 12, where the connection with type 1 is a regular one and the
connection with type 2 bypasses two extreme segments. In Figure~\ref{fig17}, both connections with type 2 bypass
one extreme segment.

\section{Conclusions} 

\begin{figure}[h!]
\centerline{\includegraphics[scale=0.6]{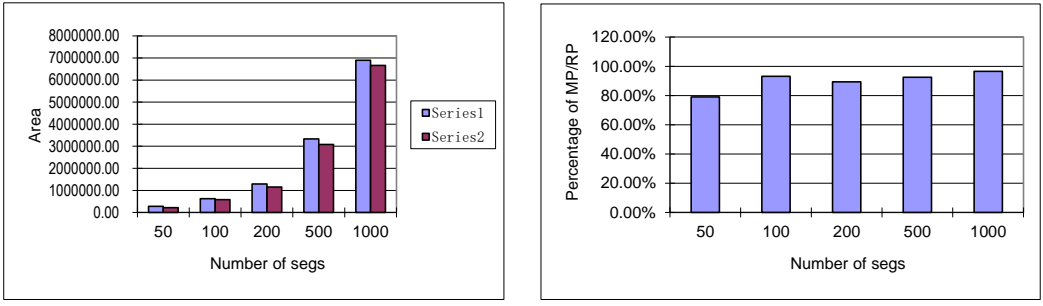}}
\vspace{-1in}
\caption{Experiments with a heuristic approach \label{fig:pic18}}
\end{figure}

It is possible that a user may not be interested in computing $P_{min}$, but rather a convex approximation that
intersects all the segments. For this, we suggest the following heuristic. Pick a point at random from each segment and
report the convex hull of these $n$ points as an approximation of the minimum area polygon. Repeating and averaging over a 
large number of runs might work out to be even better. We include a table below (see Fig.~\ref{fig:pic18}) from an experiment we made that confirms 
this intuition. For example, on a 1000 segments input, the average area of 50 convex polygon stabbers is within 96.59\% of the optimal.    \\

A more sophisticated approach might be to identify a small (constant ?) fraction of the input (a la core sets) that can be used to
obtain a provably good approximate solution. This is worth exploring further. \\  

\section*{Acknowledgements}
This research was supported by an NSERC Discovery Grant to the third author.

\end{document}